\documentclass[twocolumn]{aastex62}

\usepackage{txfonts}
\usepackage{graphicx}
\usepackage{color}

\begin{document}

\title{Saddle-shaped solar flare arcades}

\correspondingauthor{J. L\"{o}rin\v{c}\'{i}k}
\email{lorincik@asu.cas.cz}

\author[0000-0002-9690-8456]{Juraj L\"{o}rin\v{c}\'{i}k}
\affil{Astronomical Institute of the Czech Academy of Sciences, Fri\v{c}ova 298, 251 65 Ond\v{r}ejov, Czech Republic}
\affil{Institute of Astronomy, Charles University, V Hole\v{s}ovi\v{c}k\'{a}ch 2, CZ-18000 Prague 8, Czech Republic}

\author[0000-0003-1308-7427]{Jaroslav Dud\'{i}k}
\affil{Astronomical Institute of the Czech Academy of Sciences, Fri\v{c}ova 298, 251 65 Ond\v{r}ejov, Czech Republic}

\author[0000-0001-5810-1566]{Guillaume Aulanier}
\affil{LESIA, Observatoire de Paris, Universit\'e PSL , CNRS, Sorbonne Universit\'e, Universit\'e de Paris, 5 place Jules Janssen, 92190 Meudon, France}
\affil{Rosseland Centre for Solar Physics, University of Oslo, P.O. Box 1029 Blindern, NO-0315 Oslo, Norway}

\begin{abstract}

Arcades of flare loops form as a consequence of magnetic reconnection powering solar flares and eruptions. We analyse the morphology and evolution of flare arcades that formed during five well-known eruptive flares. We show that the arcades have a common saddle-like shape. The saddles occur despite the fact that the flares were of different classes (C to X), occurred in different magnetic environments, and were observed in various projections. The saddles are related to the presence of longer, relatively-higher, and inclined flare loops, consistently observed at the ends of the arcades, which we term `cantles'. Our observations indicate that cantles typically join straight portions of flare ribbons with hooked extensions of the conjugate ribbons. The origin of the cantles is investigated in stereoscopic observations of the 2011 May 9 eruptive flare carried out by the Atmospheric Imaging Assembly (AIA) and Extreme Ultraviolet Imager (EUVI). The mutual separation of the instruments led to ideal observational conditions allowing for simultaneous analysis of the evolving cantle and the underlying ribbon hook. Based on our analysis we suggest that the formation of one of the cantles can be explained by magnetic reconnection between the erupting structure and its overlying arcades. We propose that the morphology of flare arcades can provide information about the reconnection geometries in which the individual flare loops originate. 

\end{abstract}


\section{Introduction}
\label{sec_introduction}

Solar flares are sudden releases of magnetic energy accumulated in the solar atmosphere \citep[see e.g.,][]{schmieder15}. The flare emission originates in flare loops \citep[see e.g.,][]{tsuneta92} formed by magnetic reconnection, initially described in the two-dimensional `CSHKP' model of flares \citep[see e.g.,][and the references therein]{shibata11}. They are nowadays routinely observed by the Atmospheric Imaging Assembly \citep[AIA;][]{lemen12,boerner12} onboard the \textit{Solar Dynamics Observatory} \citep[\textit{SDO};][]{pesnell12}.

Arcades of flare loops develop along polarity inversion lines \citep[e.g.,][]{moore92,priest02}. Their footpoints form flare ribbons, bright elongated structures observed in lower atmosphere \citep[e.g.,][]{fletcher04}. Since both the flare loop arcades and ribbons develop along the dimension missing in the CSHKP model, 2.5D models, based on stacking the 2D models along the third dimension, have been developed \citep[e.g.,][]{shiota05, tripathy06}. These models often show translation symmetry, and are thus insufficient for description of arcades with complicated morphology. Among these can be, for example, arcades forming along curved PILs \citep[e.g.,][]{ryutkova11, baker20}, or arcades distorted by the supra-arcade downflows \citep[SADs; e.g.,][]{savage10,savage12,cassak13,xue20}. Moreover, arcades evolve in time due to ongoing reconnection and formation of new flare loops \citep{masuda01}, causing the arcades to grow \citep{schmieder95} and rise \citep{gallagher02}, see also Section 1.3 in \citet{fletcher11}. A little attention has however been paid to arcade asymmetries caused by variations of both lengths and heights of flare loops they are composed of.

\begin{figure*}[t]
  \centering    
    \includegraphics[width=6.00cm, clip,   viewport=05 17 243 327]{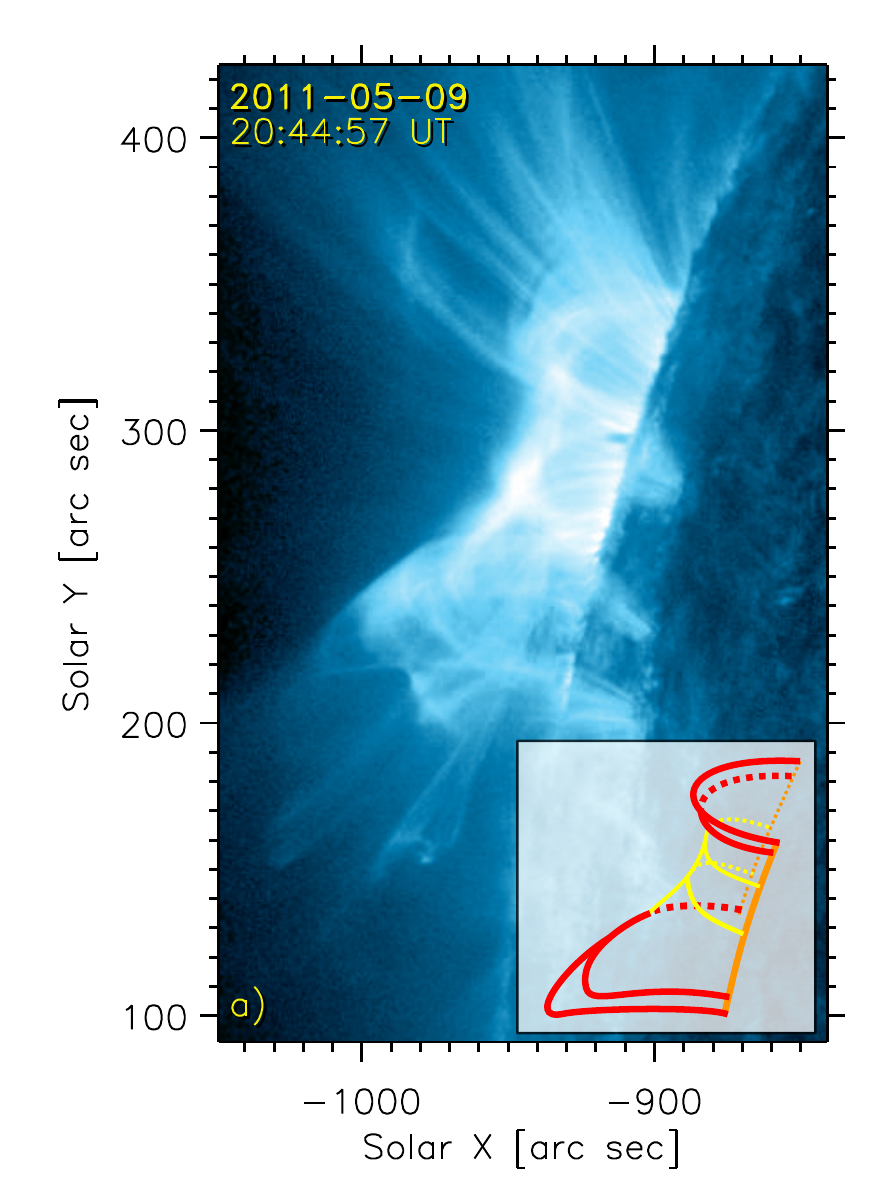}
    \includegraphics[width=5.55cm, clip,   viewport=23 17 243 327]{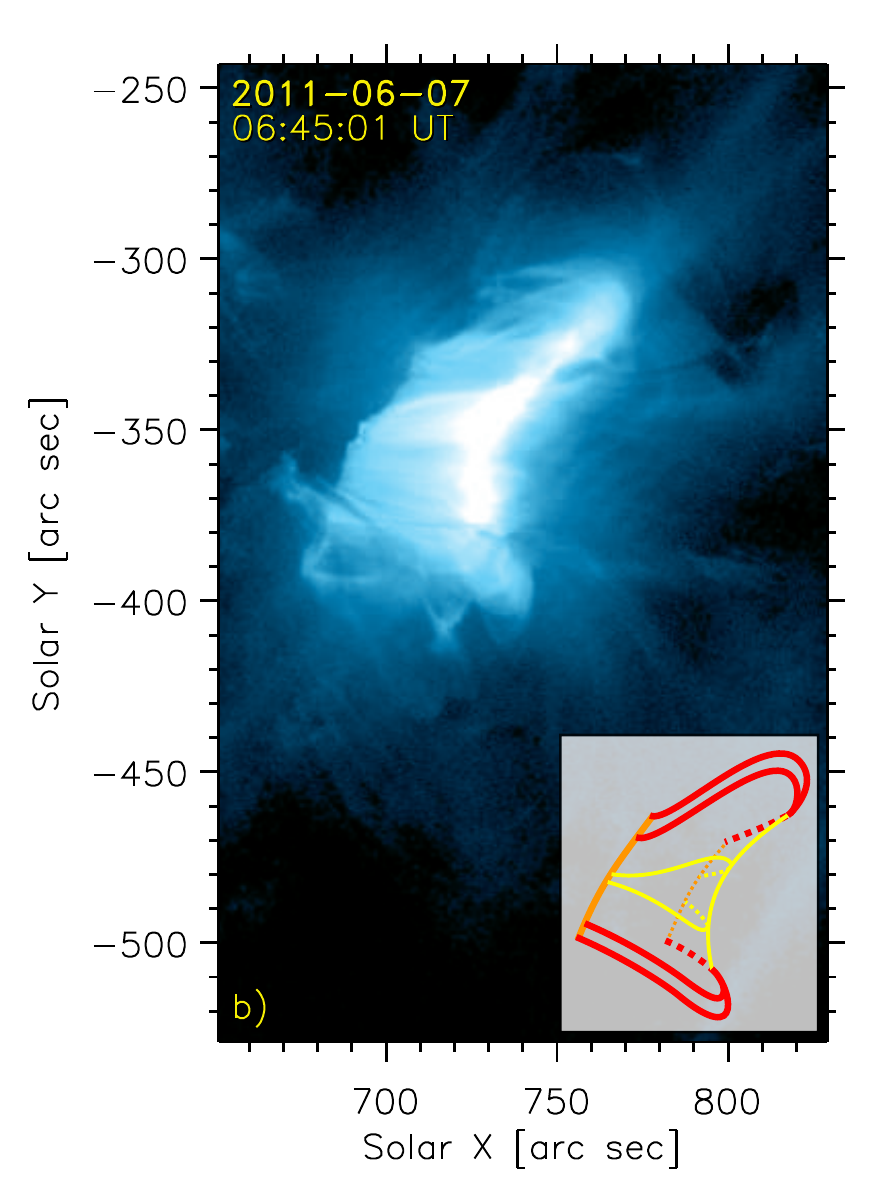}
    \includegraphics[width=5.55cm, clip,   viewport=23 17 243 327]{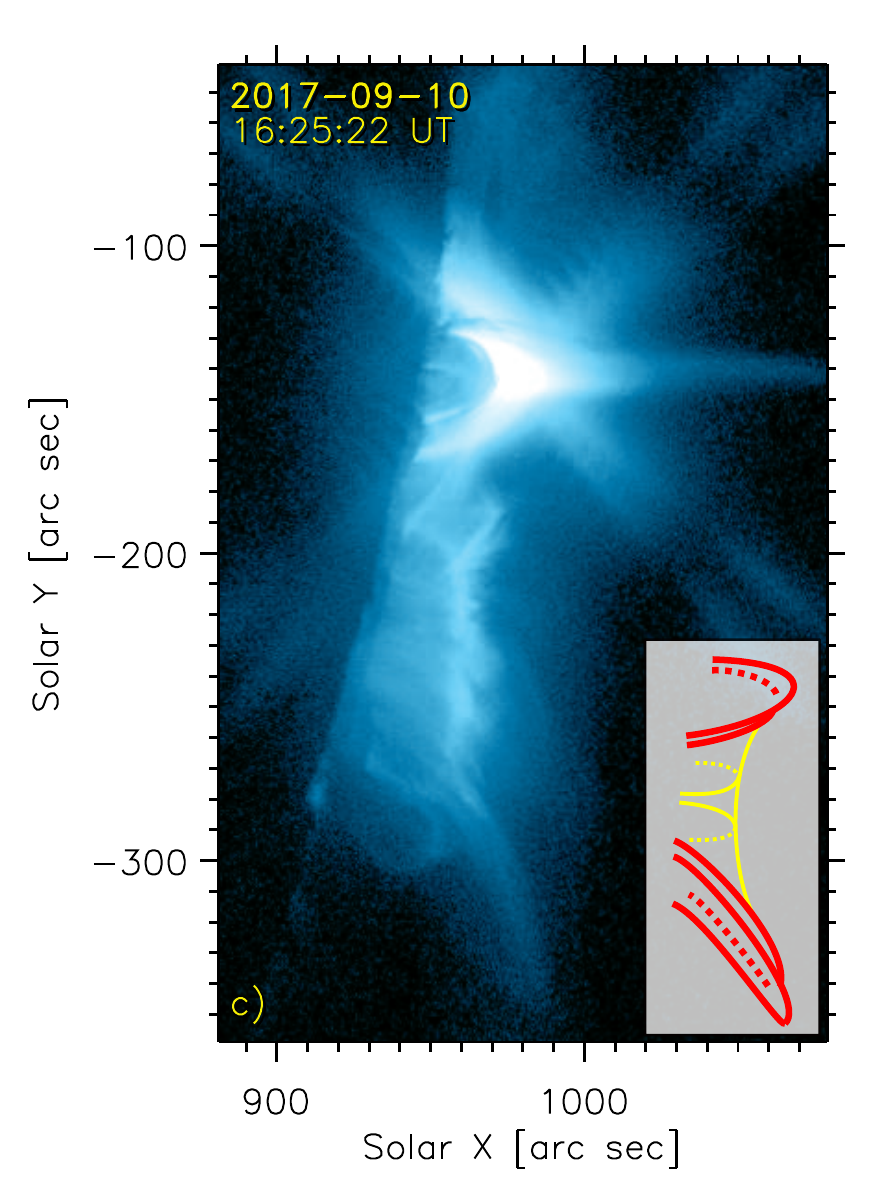}
	\\
    \includegraphics[width=8.50cm, clip,   viewport=10 25 360 295]{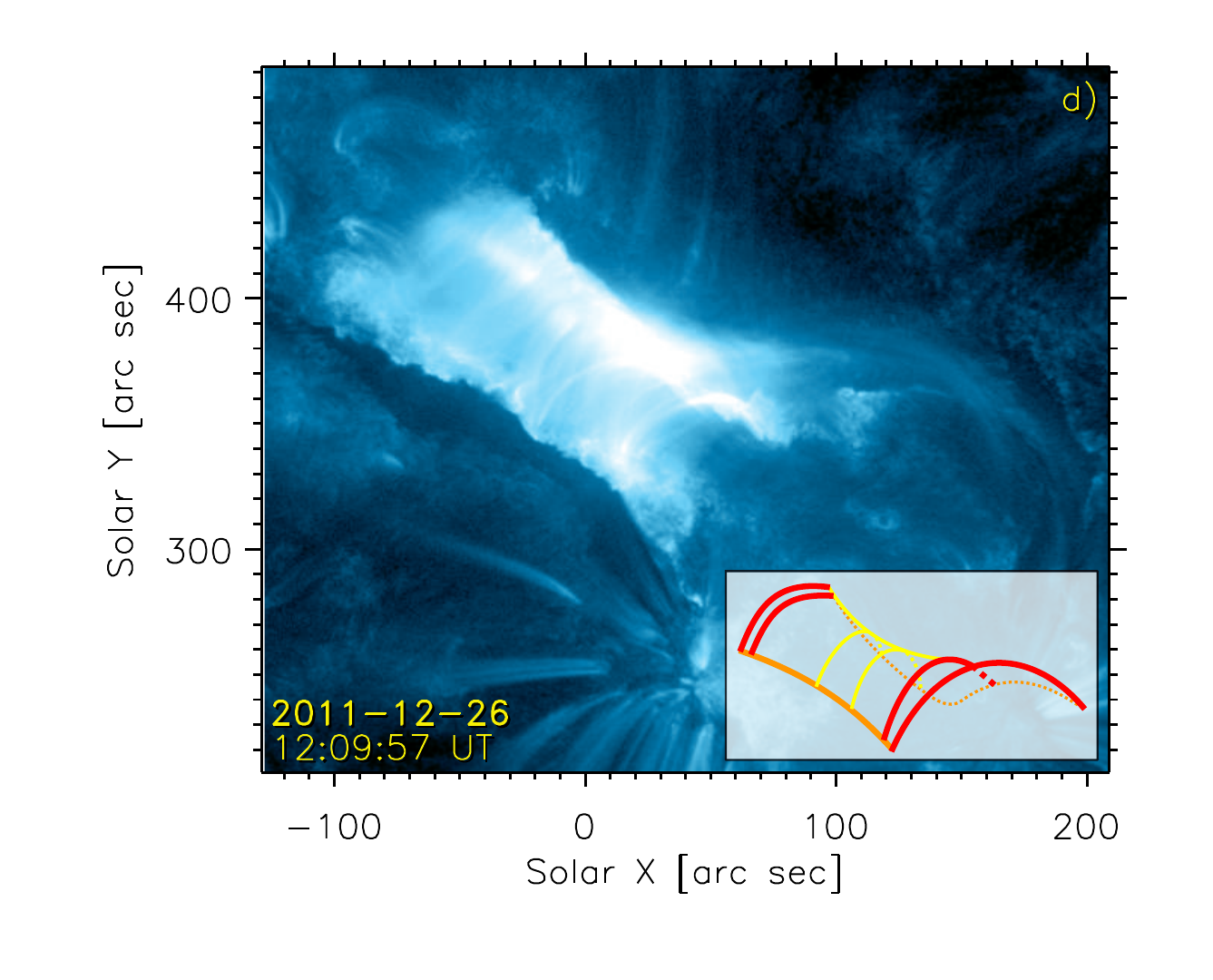}
    \includegraphics[width=7.65cm, clip,   viewport=45 25 360 295]{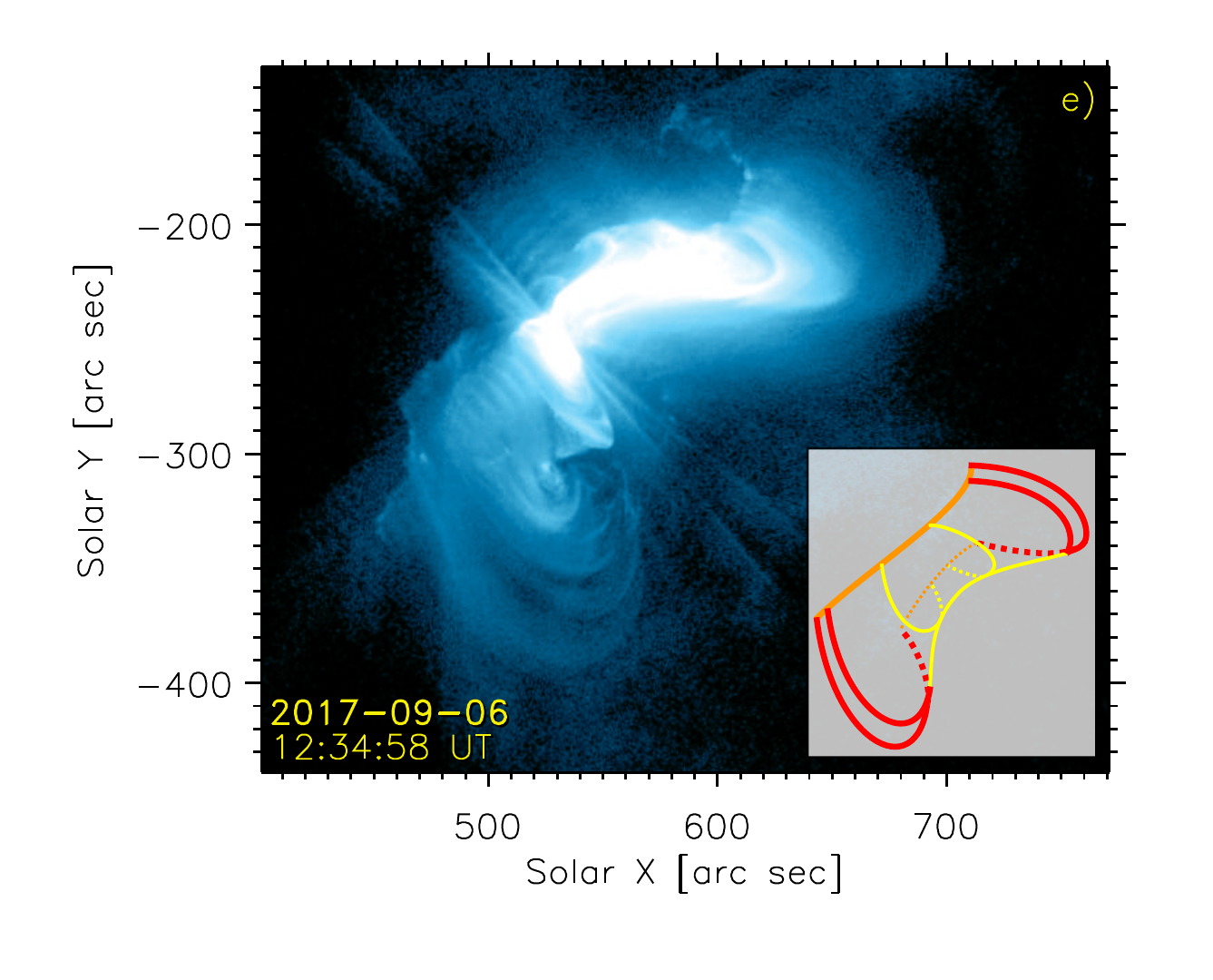}
  \caption{Saddle-shaped arcades of flare loops during five eruptive flares as observed in the 131\,\AA~filter channel of AIA. In bottom-right of each panel, cartoons showing the arcade in the observed projection is drawn. Loops at ends of the arcades are labeled as `cantles' and shown in red, while loops in central parts of the arcades are plotted in yellow. Where possible, flare ribbons are indicated using orange lines. 
  \label{fig_ov_cartoons}}
\end{figure*} 

Recent flare analyses \citep{zemanova19, lorincik19, chen19} reported on flare loops rooted in the vicinity of $J$-shaped extensions of flare ribbons called ribbon hooks (see e.g., \citet{janvier17}). Arcades reaching toward the ribbon hooks can be identified in 3D simulations of flares and eruptions \citep[e.g.,][]{inoue14, inoue15, torok18}, which suggests that their formation is associated with 3D magnetic reconnection. According to the The Standard flare model in 3D \citep{aulanier12, janvier13}, flare loops rooted in the hooks should form due to the $\textit{ar--rf}$ reconnection between the erupting structure and its overlying arcades \citep{aulanier19, dudik19}. This could introduce additional asymmetries to flare arcades. 

In this manuscript we analyse the evolution and morphology of saddle-shaped flare arcades formed during five events well-known from literature. Section \ref{sec_data} briefly introduces the data used in this study. In Sections \ref{sec_110509} and \ref{sec_other} we analyse flare arcades formed during the five events. Finally, Section \ref{sec_summary} summarizes our results. 

\begin{figure*}[!t]
  \centering    
  	\includegraphics[width=18.00cm, clip,  viewport=   0 20 680 465]{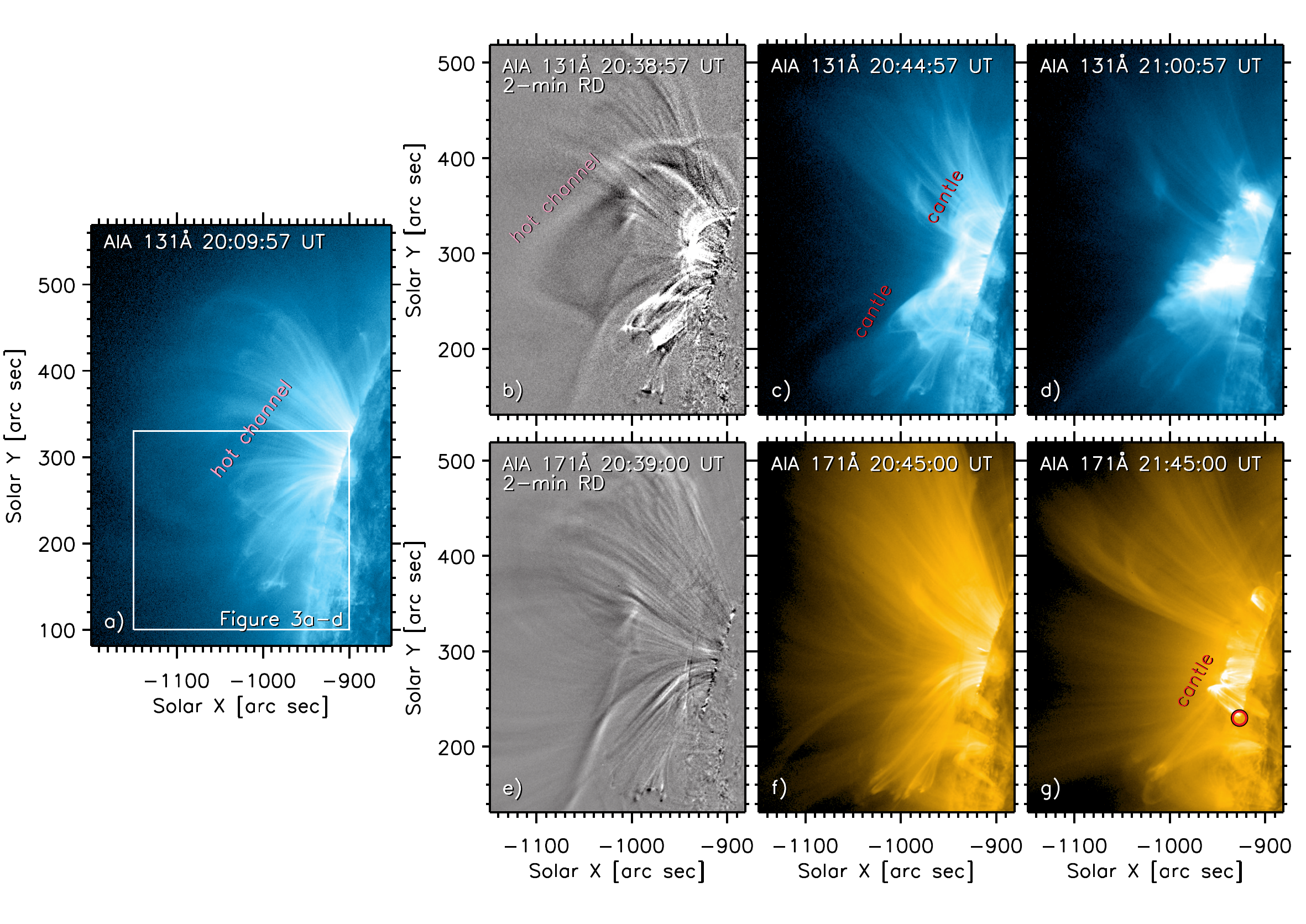}
  	\\  	
    \includegraphics[width=9.00cm, clip,   viewport=  10 0 273 190]{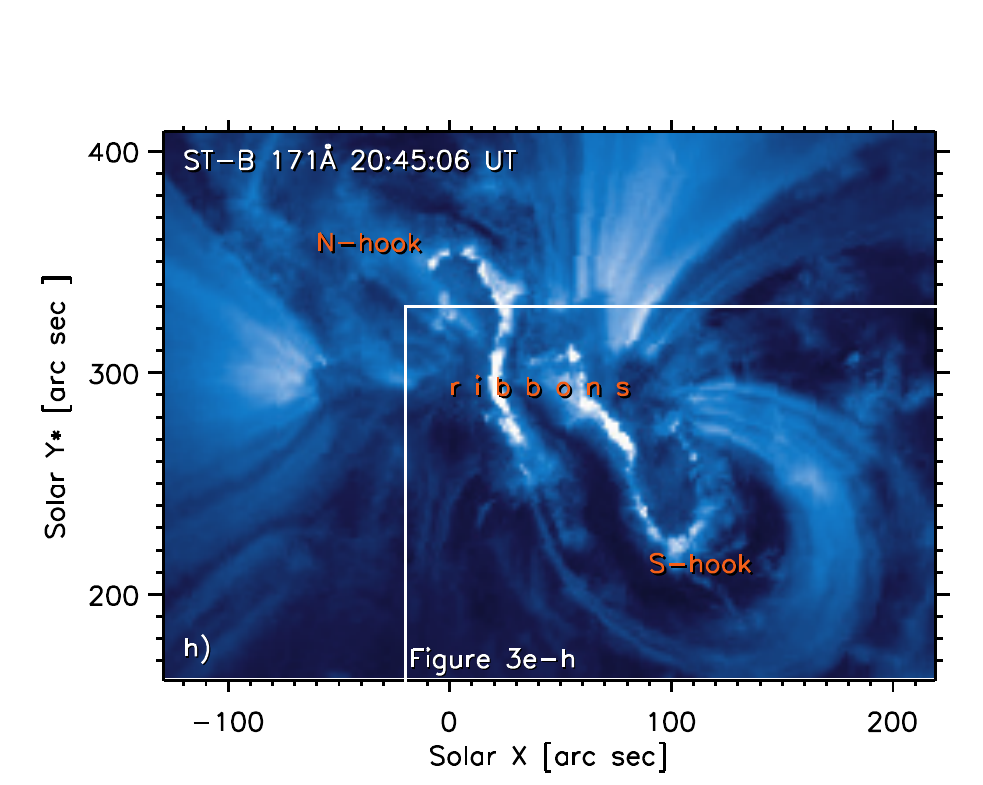}
    \includegraphics[width=7.80cm, clip,   viewport=  45 0 273 190]{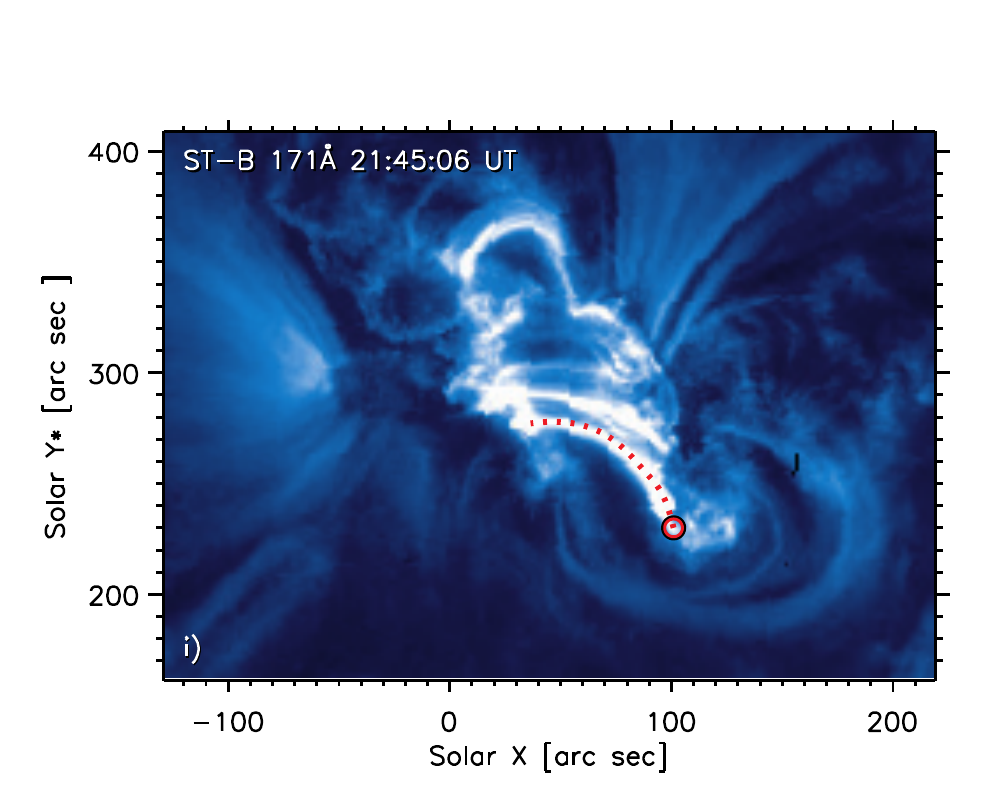}
  \caption{The 2011 May 9 eruption observed in the 131\,\AA~and 171\,\AA~channels of AIA (panels (a)--(g)) and the 171\,\AA~channel of \textit{STEREO-B}/EUVI (panels (h)--(i)). The hot channel before and during its eruption, cantles of the saddle-shaped arcade of flare loops, and hooked flare ribbons are indicated. Red circle plotted in panels (g) and (i) marks the southern footpoint of the southermost flare loop (cantle; indicated using red dotted line in panel (i). \\ Animated version of the observations from the 131\,\AA~and 171\,\AA~channels of AIA and 171\,\AA~channel of EUVI is available online. The animation contains the eruption of the hot channel, evolution of the surrounding corona, and the formation of the hooked flare ribbons. It covers the period 20:00 -- 22:00 UT and its real-time duration is 30 seconds.  \label{fig_110509}}
\end{figure*}
\section{Data and observations}
\label{sec_data}

The events discussed in this manuscript are primarily analysed using imaging data from \textit{SDO}/AIA. AIA provides full-disk images of the Sun in 10 filters, imaging plasma with temperatures between $\approx$10$^4$ to 10$^7$ K. The spatial resolution of AIA is $\approx$1.5\arcsec \citep{boerner12} and its cadence is 12--24\,s depending on bandpass. We processed level-1 AIA data using the \texttt{aia\_prep} routine. Flares located on solar disk were corrected for the differential rotation.

Where appropriate, we supplement AIA data with observations performed by the Extreme Ultraviolet Imager \citep[EUVI;][]{wuesler04, howard08} onboard STEREO-B, as well as the X-Ray Telescope \citep[XRT;][]{golub07} onboard the \textit{Hinode} mission. These datasets were processed using standard routines available within SolarSoft and co-registered with AIA. These shifted coordinates are denoted as Solar-$X^*$ and Solar-$Y^*$. Finally, the phases of individual flares were analysed using the soft X-ray flux measured by \textit{GOES}.

We chose eruptive flares of different classes (C to X) known from literature, all showing well-developed arcades and hooked ribbons, with the exception of one event seen off-limb where the ribbons could not be observed (Section \ref{sec_20170910}). Figure \ref{fig_ov_cartoons} provides an overview of the morphology of flare arcades formed during the selected events, all observed in the 131\,\AA~channel of AIA. Each individual panel also contains a cartoon showing the simplified arcade morphology as observed. All of these arcades are asymmetric along the identifiable ribbons (orange lines) to some degree. Flare loops at each end of the arcade (red) are inclined and relatively-higher than the central parts of the arcades (yellow). Overall, all five arcades are reminiscent of saddles. We label the inclined, higher loops at each end of each arcade as `cantle' (red). Note that the saddles are also observed in flares which have nearly parallel ribbons (and thus straight PILs), see panels (a), (b), and (d) of Figure \ref{fig_ov_cartoons}. To our knowledge, this overall saddle-shaped morphology of the flare arcades was not noticed so far. We now proceed to detail the evolution and morphology of these arcades. 

\begin{figure*}[t]
  \centering    
    \includegraphics[width=5.15cm, clip,   viewport=  20 05 258 220]{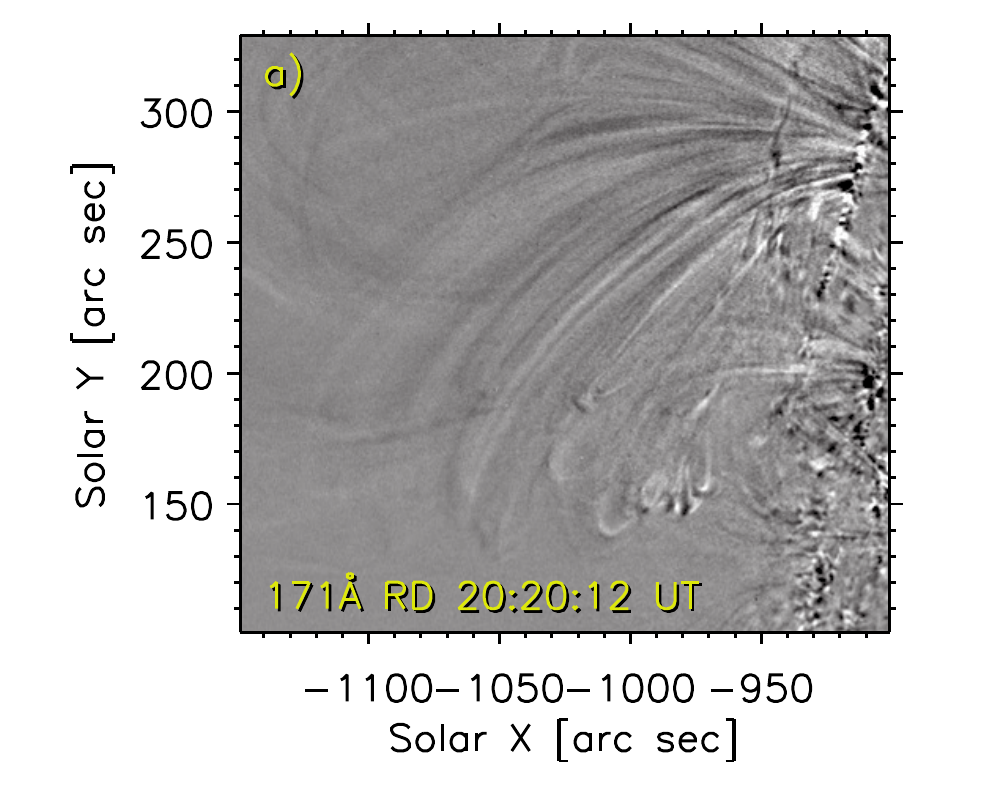}
    \includegraphics[width=4.13cm, clip,   viewport=  67 05 258 220]{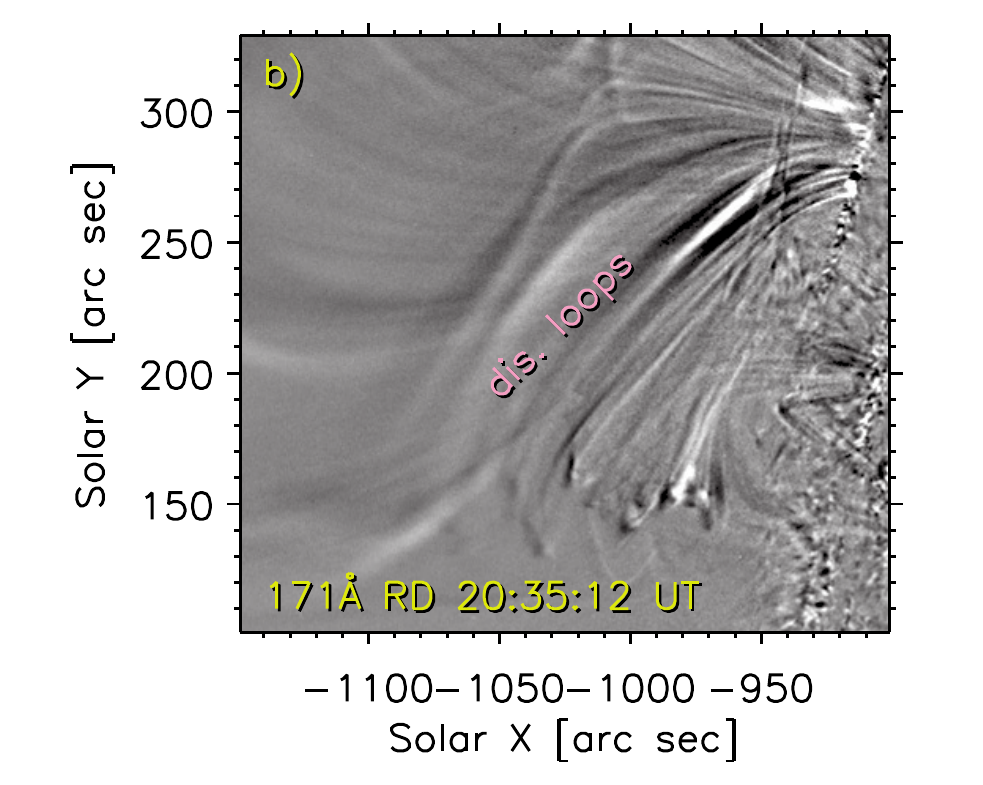}
    \includegraphics[width=4.13cm, clip,   viewport=  67 05 258 220]{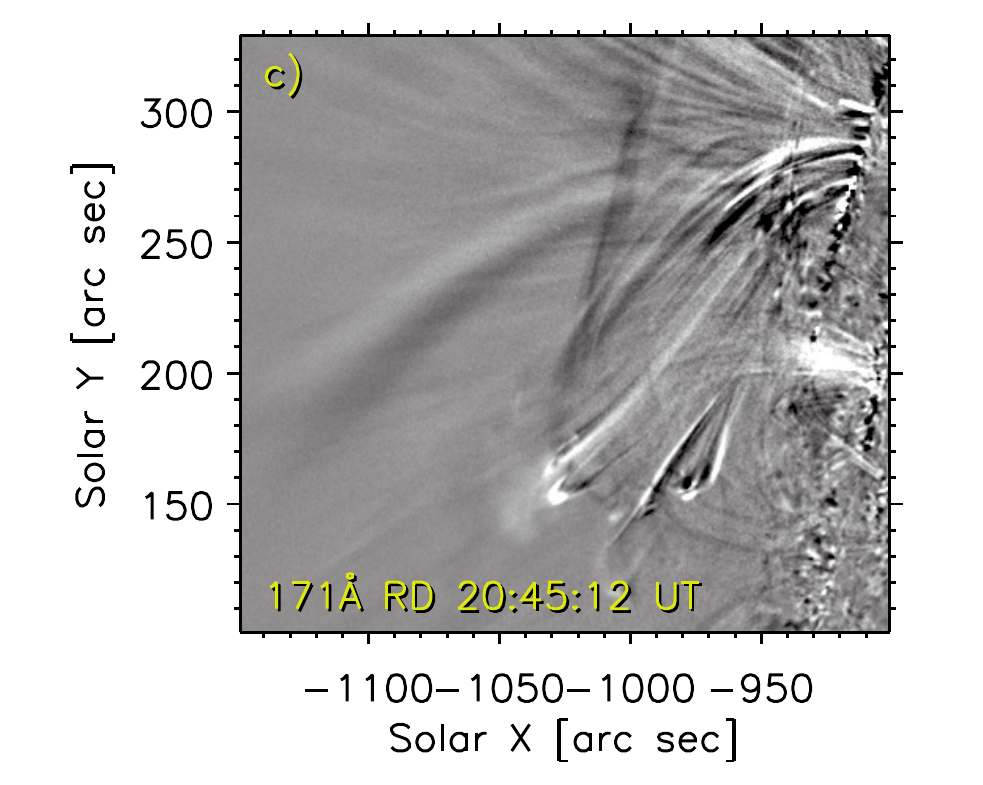}
    \includegraphics[width=4.13cm, clip,   viewport=  67 05 258 220]{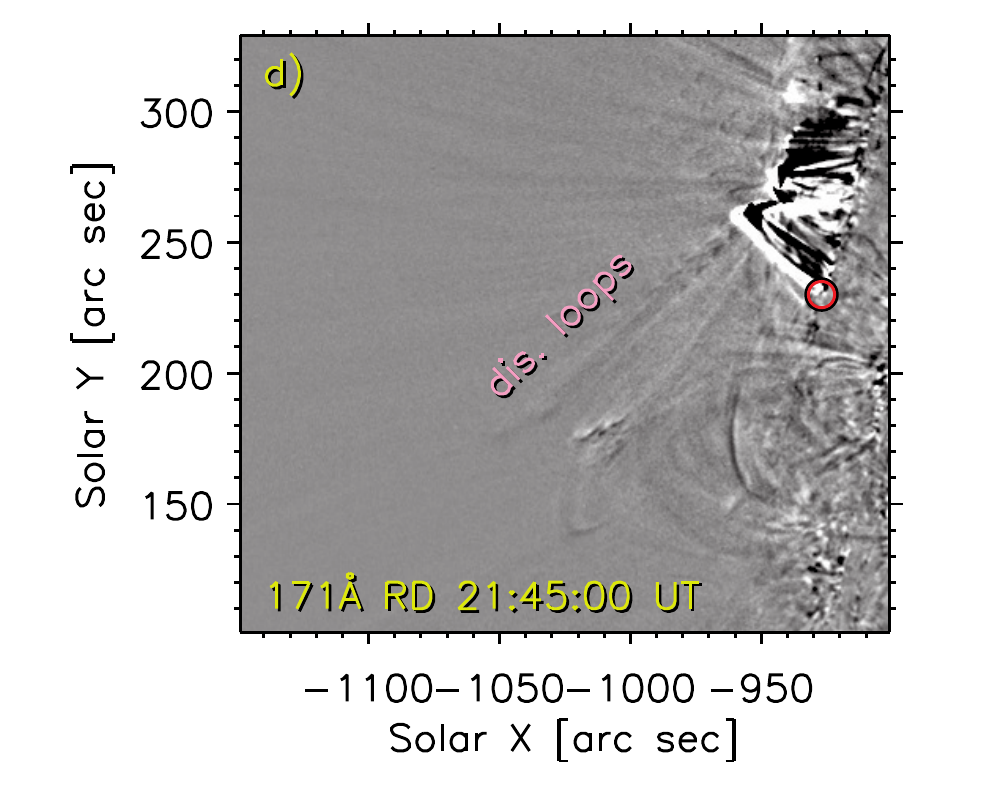}
    \\
    \includegraphics[width=5.15cm, clip,   viewport=  20 05 258 220]{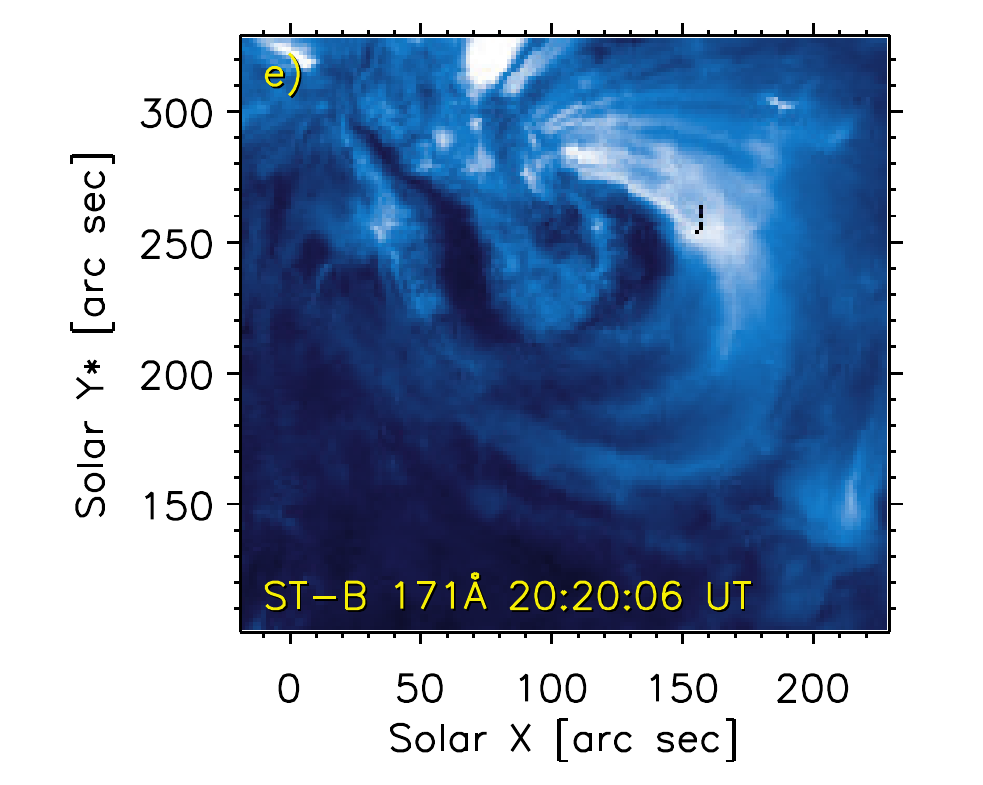}
    \includegraphics[width=4.13cm, clip,   viewport=  67 05 258 220]{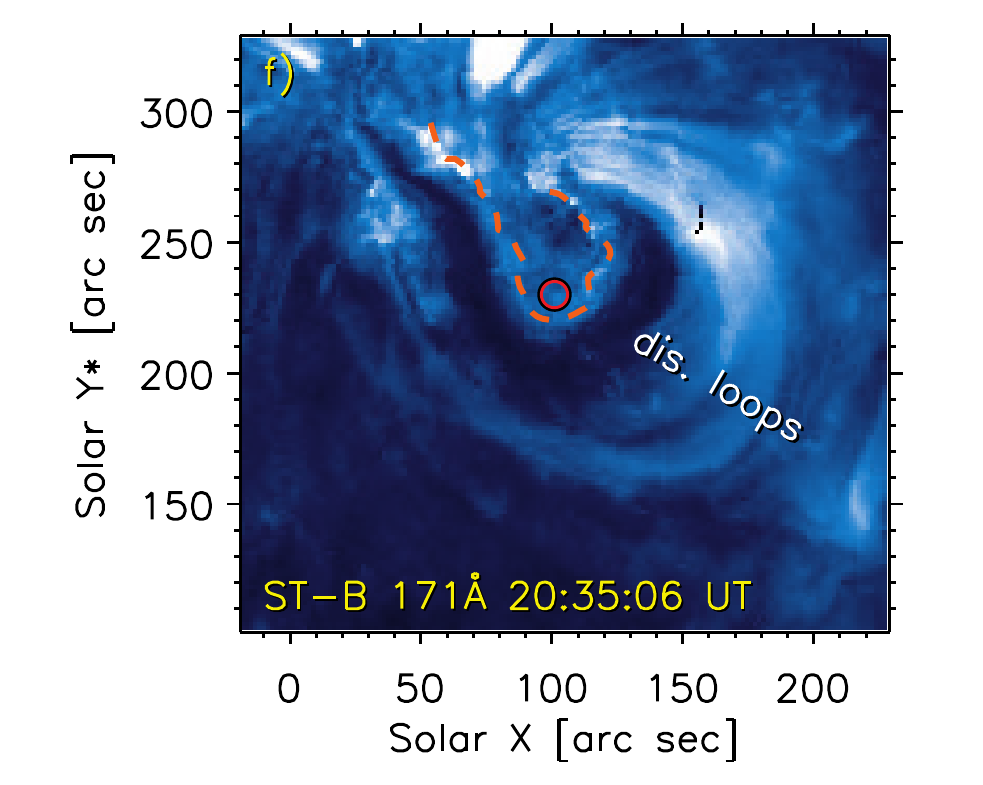}
    \includegraphics[width=4.13cm, clip,   viewport=  67 05 258 220]{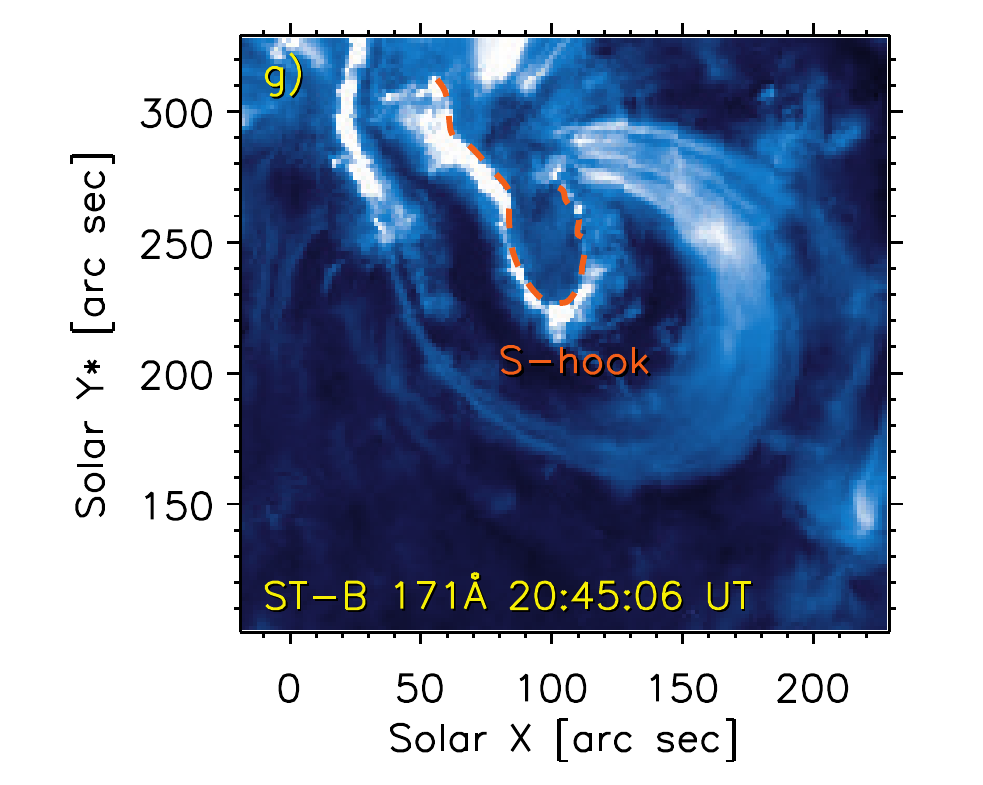}
    \includegraphics[width=4.13cm, clip,   viewport=  67 05 258 220]{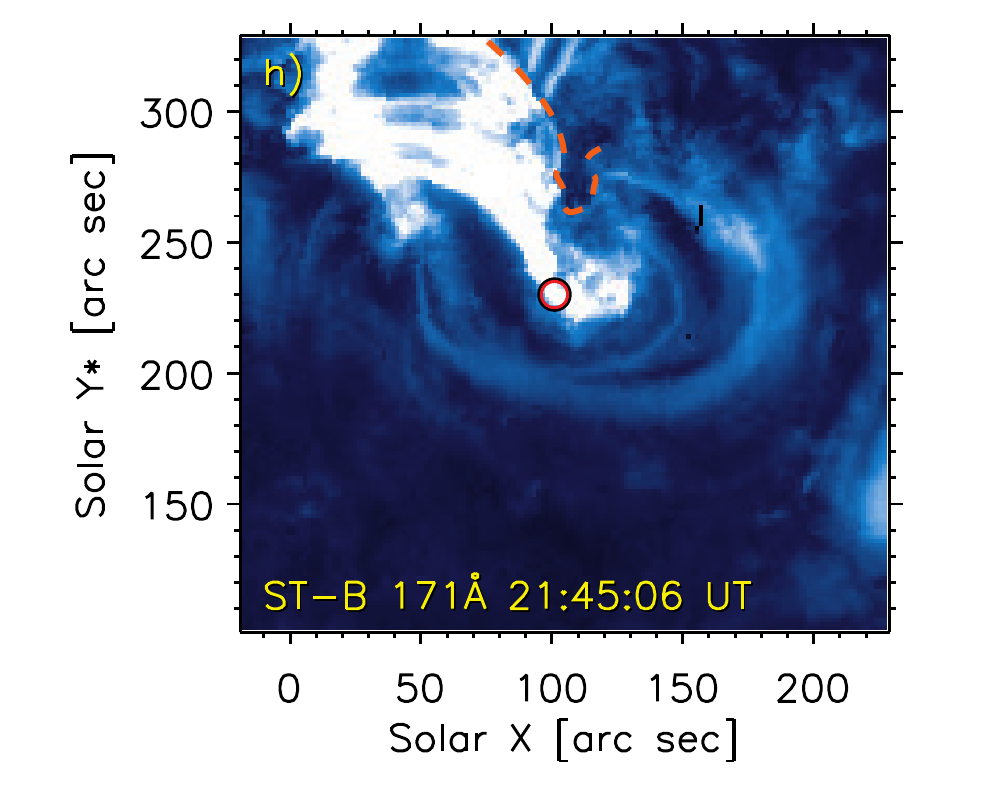}
  \caption{View of the disappearing coronal loops in 2-minute running-difference images produced using the 171\,\AA~filter channel of AIA (panels (a)--(d)) and in STEREO-B 171\,\AA~channel (panels (e)--(h)). In panels (f)--(h), the hooked flare ribbon is indicated using orange dashed line. Red circle in panels (d), (f), and (h) marks the southern footpoint of the southermost flare loop. \label{fig_110509_rec}}
\end{figure*}
\section{Saddle during the 2011 May 9 eruption}
\label{sec_110509}

We first analyze the 2011 May 9 eruption accompanied by a C5.4-class flare. This event is ideal for our investigation as it was observed stereoscopically by AIA and EUVI with angular separation of about 94$^{\circ}$. AIA observed it at the limb, while in EUVI it was near the disk center.
\subsection{Arcade of flare loops}
\label{sect_110509_ov}

The erupting structure was a hot channel \citep[e.g.,][]{zhang12} visible in the 94\,\AA~and 131\,\AA~channels of AIA. Before its eruption, the hot channel was embedded in an arcade of coronal loops (Figure \ref{fig_110509}(a)--(b)) within a bipolar active region NOAA 11193 \citep{aulanier12}. The hot channel started to slowly rise after $\approx$20:30 UT. In  2-minute running-difference (RD) 131\,\AA~images, constructed from images averaged over 1 minute, both the rising hot channel and the first flare loops can be distinguished (panel (b)). During the impulsive phase of the flare at $\approx$20:45 UT, a completely-formed arcade of flare loops is seen in the 131\,\AA~channel (panel (c)). It has a shape of a saddle; the flare loops composing its far ends are relatively-higher compared to those in its central part. The high flare loops at the south are also inclined. By analogy with the horse-back riding terminology, we refer to these lifted ends of saddles as `cantles'. The saddle is visible for about fifteen minutes. During the flare peak (panel (d)), the saddle starts to be obscured by the developing supra-arcade downflows \citep[see][]{warren11}. The southern cantle can however be discerned later in filters such as 171\,\AA~(panel (g)), or 193\,\AA~and 211\,\AA, a consequence of cooling of flare loops. 

At this stage, some of the flare loops forming the cooling arcade can also be identified on-disk in the 171\,\AA~channel of EUVI (panel (i)). The lengths of these flare loops vary along the arcade. Notably, one of the footpoints (red circle) of the southermost loop (red dotted line) is located further toward the south below the curved part of the ribbon hook (Section \ref{sect_110509_cantle}), and appears more sheared. Its length, measured using circumference of an ellipse passing through its footpoints and apex, was found to be $\approx$37 Mm, while the typical length of the flare loops elsewhere in the arcade range between about 22 and 27 Mm. The cantle loop can easily be identified in AIA, where it composes the cantle at the southern end of the saddle, as it is still the southernmost loop (red circle in panel (g)). In the AIA projection, this cantle loop seen in 171\,\AA~is also relatively-higher than the remainder of the flare arcade. The relation of this cantle loop to the flare ribbons is outlined in the following section.
\subsection{Flare ribbons and coronal loops}
\label{sect_110509_cantle}

During the eruption, the coronal loops overlying the hot channel moved aside from the erupting hot channel toward the south and north (Figure \ref{fig_110509} (e)--(f)). While the northern coronal loops remained visible during the eruption, obscuring the northern cantle, most of the southern coronal loops disappeared. EUVI 171\,\AA~observations reveal that these coronal loops were rooted at either sides of the $J$-shaped (hooked) flare ribbons. The straight parts of the ribbons were nearly parallel, separated by a straight PIL \citep[see also Figure 1 in][]{aulanier12}. The ribbon hooks, one to the north-east (N-hook) and the other to the south-west (S-hook), are indicated in Figure \ref{fig_110509}(h). 

A magnified view of the coronal loops overlying the S-hook is shown in Figure \ref{fig_110509_rec}. Panels (a)--(d) contain 2-minute running-difference images produced using AIA 171\,\AA. Initially, the coronal loops with various inclinations (a) were overlying the hot channel (seen only in 131\,\AA, Figure \ref{fig_110509}(b)). At $\approx$20:35 UT (Figure \ref{fig_110509_rec}(b)), they moved toward the south. Subsequently, the coronal loops started to vanish (panel (c)) and this process continued until most of them disappeared (d). 

Footpoints of these coronal loops can be identified in EUVI 171\,\AA~observations (Figure \ref{fig_110509_rec}(e)). After the onset of the eruption at $\approx$20:35 UT, the S-hook started to be visible (orange dashed line in panel (f)), encircling a core dimming region. Afterwards, the S-hook started to shrink and propagate toward the north (g--h). Its tip then elongated toward the north and swept the footpoints of the coronal loops. Meanwhile, these loops were disappearing and the cantle was being formed (Figure \ref{fig_110509}(c)). 

The location of the red circle originally corresponded to the region encircled by the S-hook (Figure \ref{fig_110509_rec}(f)). Later, as the hook propagated toward the north, it swept through the location of the circle, which subsequently corresponded to a flare loop footpoint (Figure \ref{fig_110509_rec}(h)). As shown by \citet{zemanova19}, \citet{lorincik19}, and \citet{dudik19} for other events, this behavior represents the change from flux rope to flare loop footpoint and is consistent with the ar--rf reconnection between the erupting flux rope and its overlying arcades \citep[see][]{aulanier19}. Furthermore, as described in Section \ref{sect_110509_ov}, the 171\,\AA~flare loop rooted in the location of the red circle is longer and relatively-higher than the remainder of the flare arcade. This behavior is also consistent with the ar--rf reconenction. Comparison of Figures 4d and 5d of \citet{aulanier19} reveals that the flare loops originating in the ar--rf geometry can be longer and also higher than the flare loops formed due to the reconnection between overlying arcades, which occurs beneath the erupting structure. A similar situation is also seen in Figure 1c therein. The formation of this cantle may therefore be explained by magnetic reconnection between the hot channel and coronal loops rooted nearby.
\section{Saddles in four additional events}
\label{sec_other}

We now show that saddle arcades also occurred in different events studied previously by numerous authors.

\begin{figure*}[t]
  \centering     
   \includegraphics[width=6.60cm, clip,   viewport=  05 0 240 270]{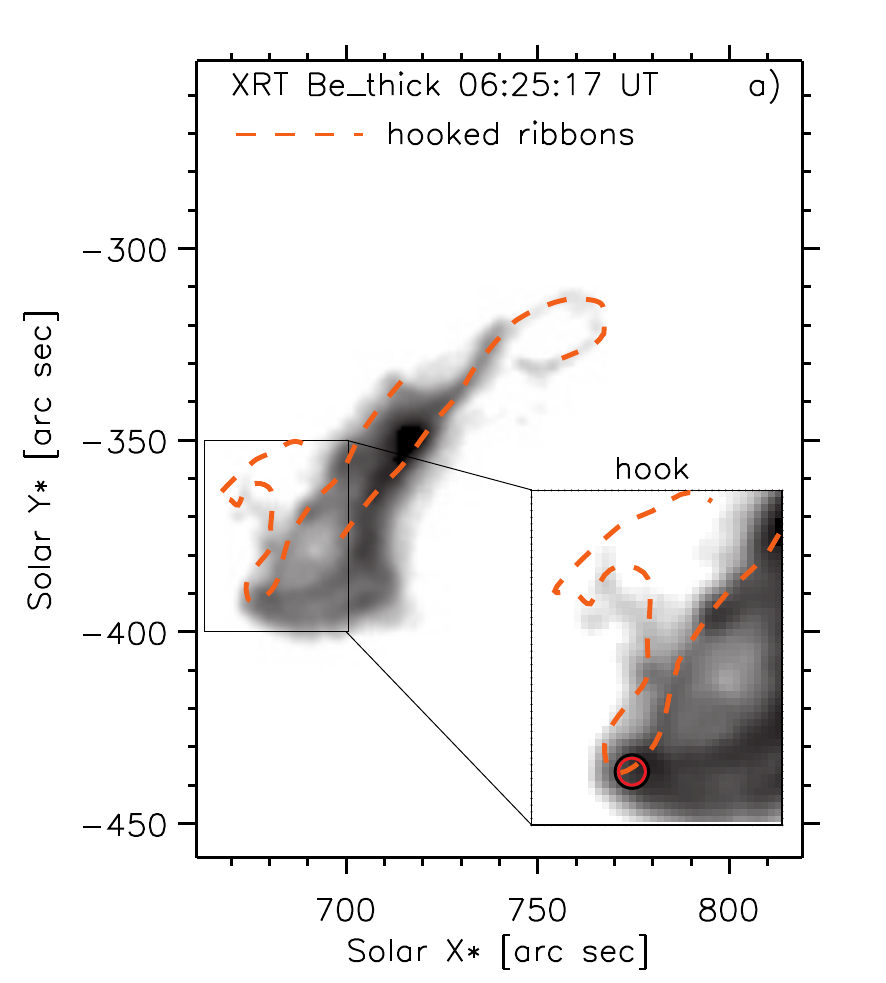}
   \includegraphics[width=5.28cm, clip,   viewport=  52 0 240 270]{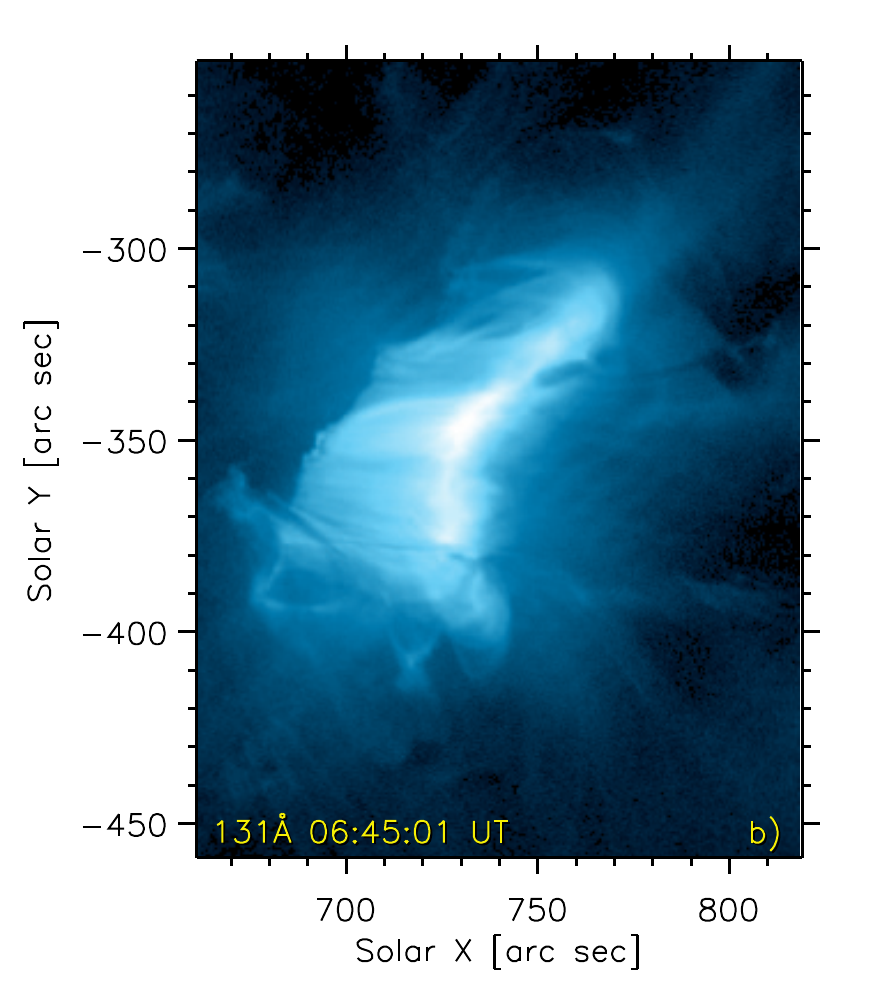}
   \includegraphics[width=5.28cm, clip,   viewport=  52 0 240 270]{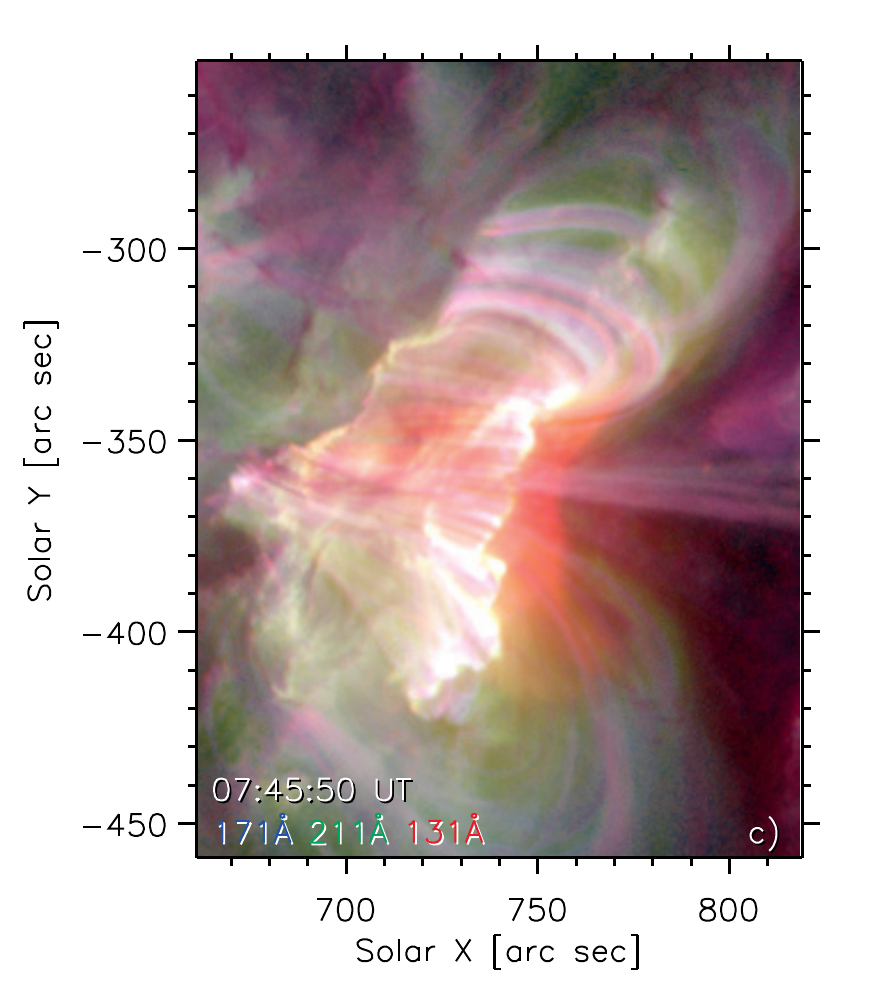}
   \\
   \includegraphics[width=6.60cm, clip,   viewport=  15 05 256 210]{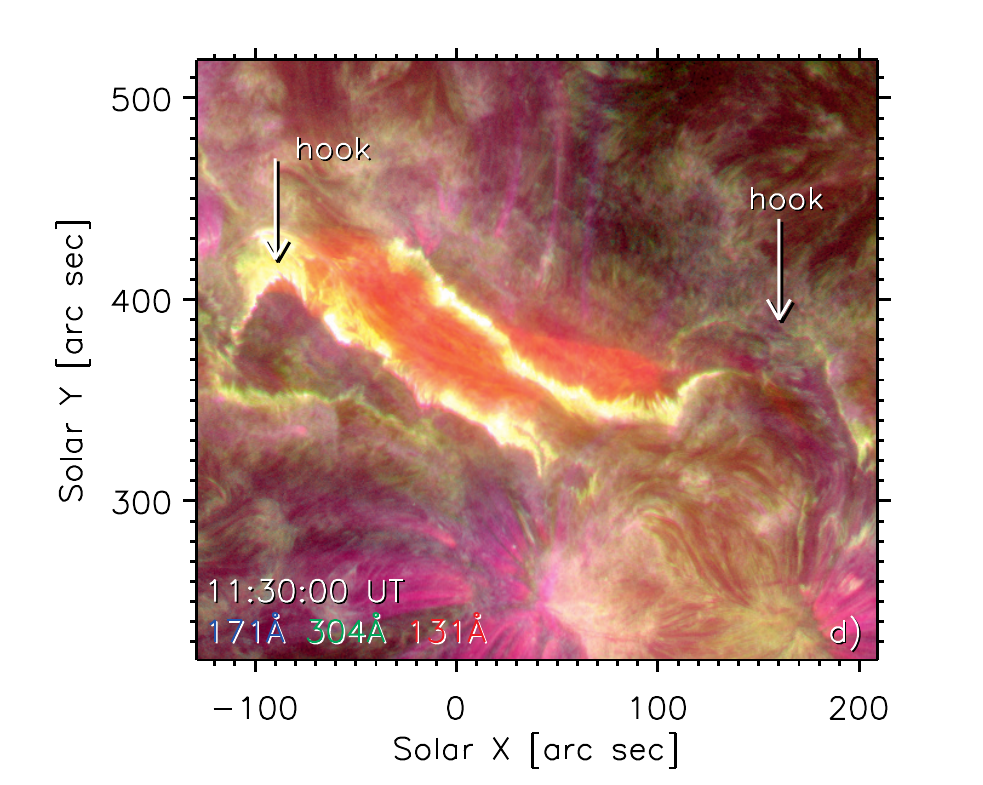}
   \includegraphics[width=5.56cm, clip,   viewport=  53 05 256 210]{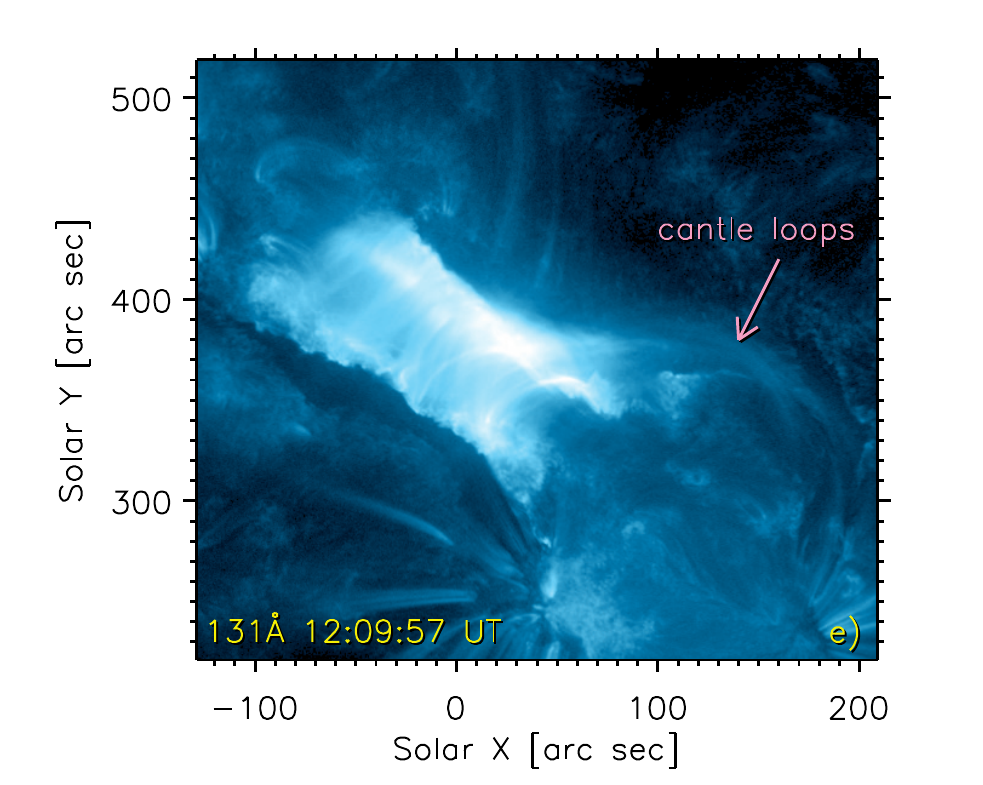}
   \includegraphics[width=5.56cm, clip,   viewport=  53 05 256 210]{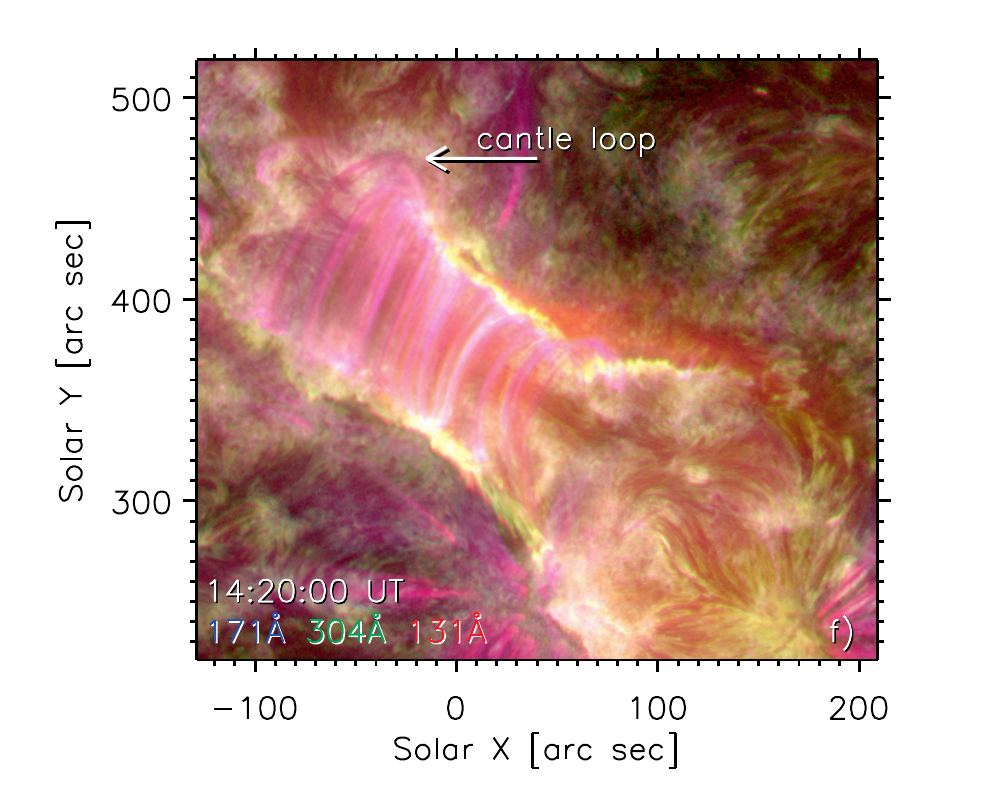}
  \caption{Saddle arcades of the 2011 June 7 (top row) and 2011 December 26 (bottom row) flares observed by XRT and AIA. Orange dashed lines plotted in panel (a) indicate the ribbon hooks of 2011 June 7 flare. The red circle marks the footpoints of cantle flare loops. Hooked ribbons and cantle loops of the 2011 December 26 flare are highlighted using arrows. \label{fig_dudik_qiu}}
\end{figure*}
\subsection{Filament eruption of 2011 June 7}
\label{sec_20110607}

The 2011 June 7 filament eruption is a well-known event \citep[e.g,][and references therein]{vanDriel14,Yardley16,dudik19}. It occurred in an active region NOAA 11226 and was accompanied by a M2.5--class flare.

This eruption started at $\approx$06:08 UT \citep{dudik19} and the first flare loops were visible roughly after 06:20 UT (first row of Figure \ref{fig_dudik_qiu}). In the 94 and 131\,\AA~channels of AIA, flare loops composing the arcade were partially obscured by the erupting filament. Early on, the arcade is however well visible in the Be\_thick filter of XRT (panel (a)). The flare loops rooted at the southern end of the arcade are longer than others, forming a cantle. About 20 minutes later (panel (b)), a full saddle can be distinguished in AIA 131\,\AA. Both cantles are visible, one inclined toward the south and the other toward the north. The saddle can be observed for more than an hour; at these later times also in cooler filter channels. The composite image shown in panel (c), produced using the 131\,\AA~(red), 171\,\AA~(blue), and 211\,\AA~(green) channel images, reveals individual flare loops composing the arcade. At many locations, emission is seen in multiple filter channels (white color).

Hooked ribbons, as traced out using AIA 1600\,\AA, are shown in \ref{fig_dudik_qiu}(a). The footpoints of flare loops composing the southern cantle, marked using the red circle in the zoomed inclusion, are located at the southern end of the ribbon hook. \citet{dudik19} shown that flare loops rooted in this region originated via the ar--rf reconnection geometry.
\begin{figure*}[t]
  \centering    
    \includegraphics[width=6.70cm, clip,   viewport=  08 10 268 250]{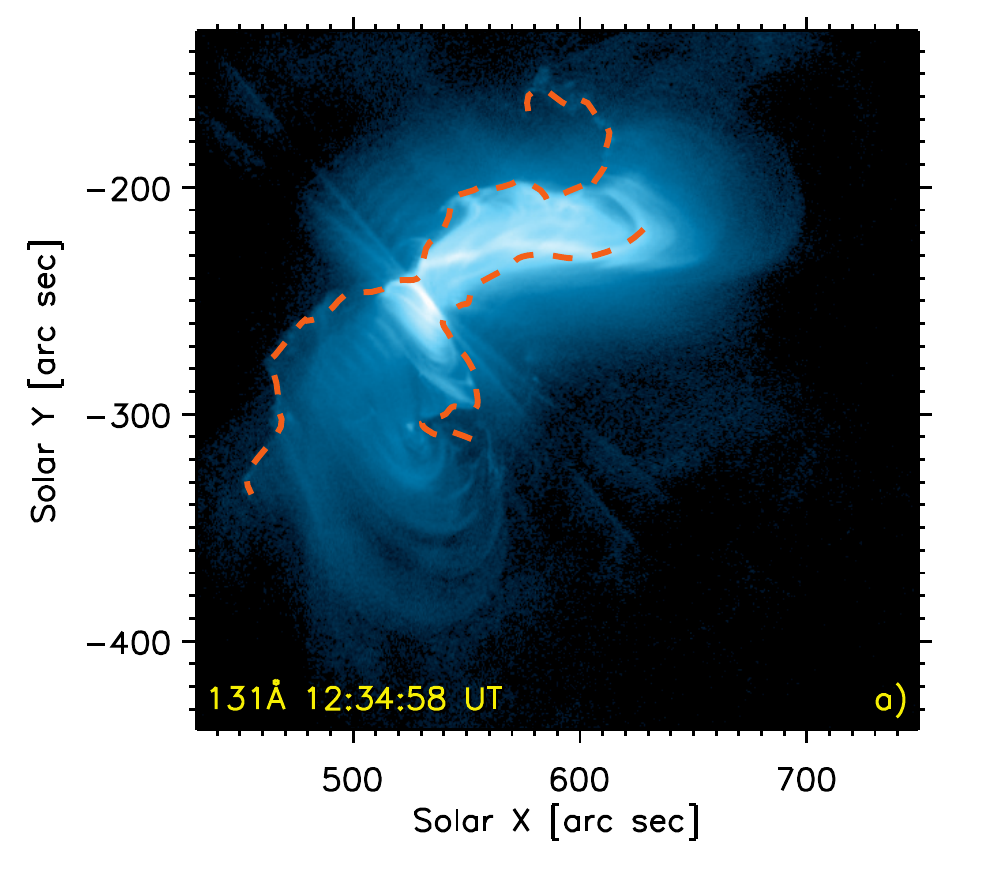}
    \includegraphics[width=5.54cm, clip,   viewport=  53 10 268 250]{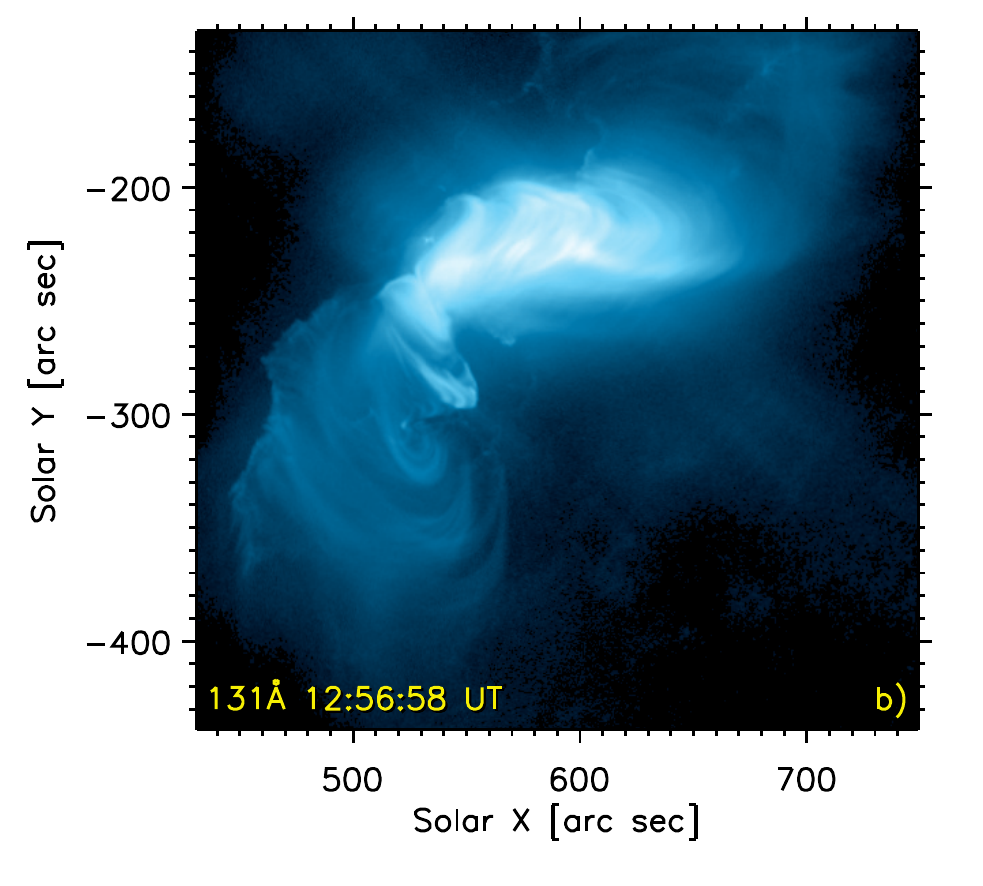}
  	\includegraphics[width=5.54cm, clip,   viewport=  53 10 268 250]{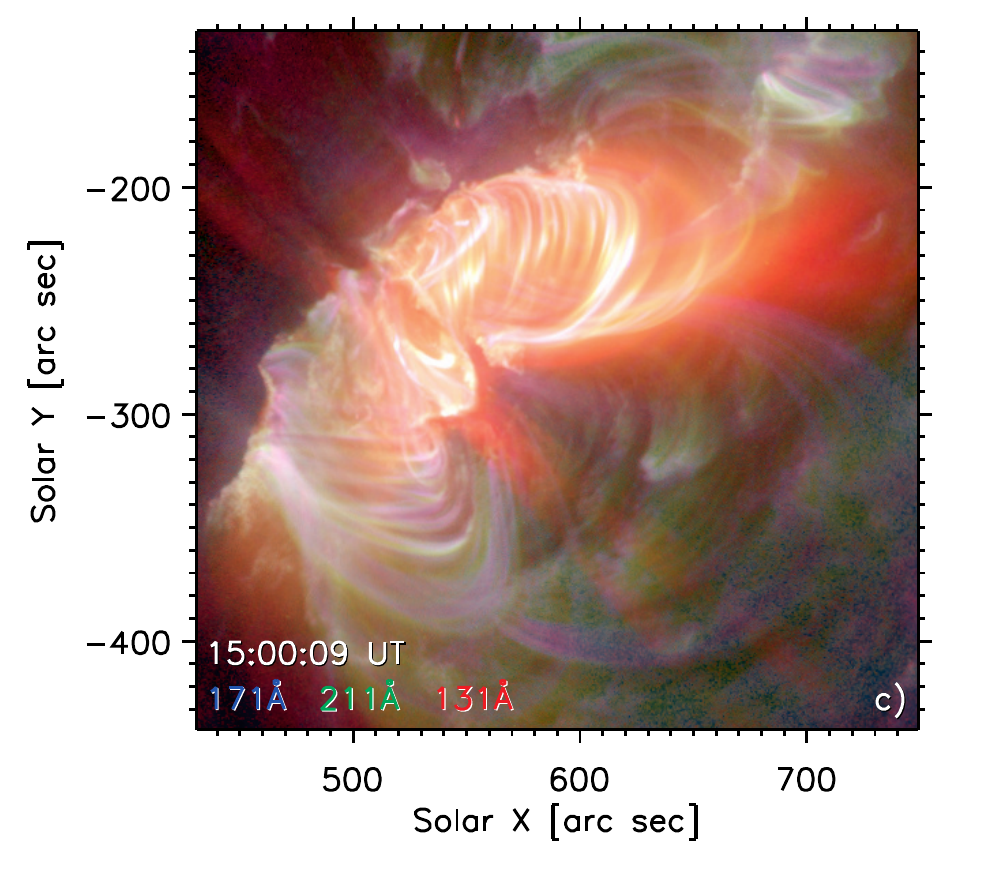}
	\\
   \includegraphics[width=4.70cm, clip,   viewport=  00 0 133 240]{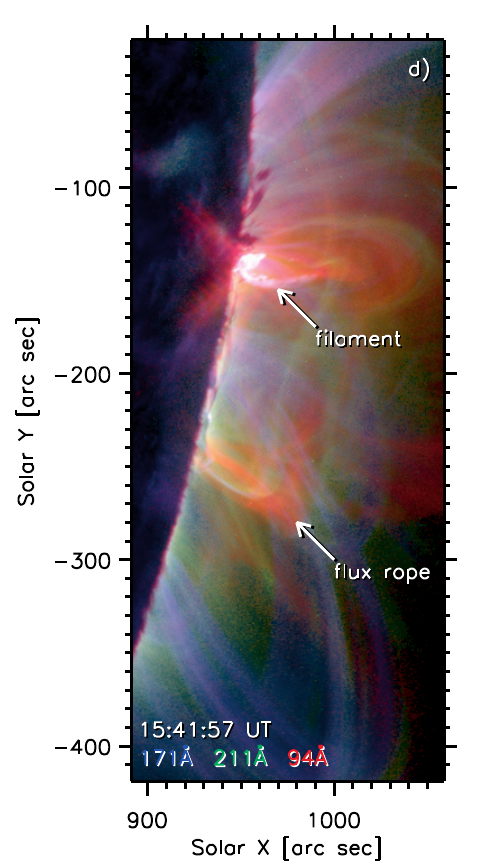}
   \includegraphics[width=3.53cm, clip,   viewport=  33 0 133 240]{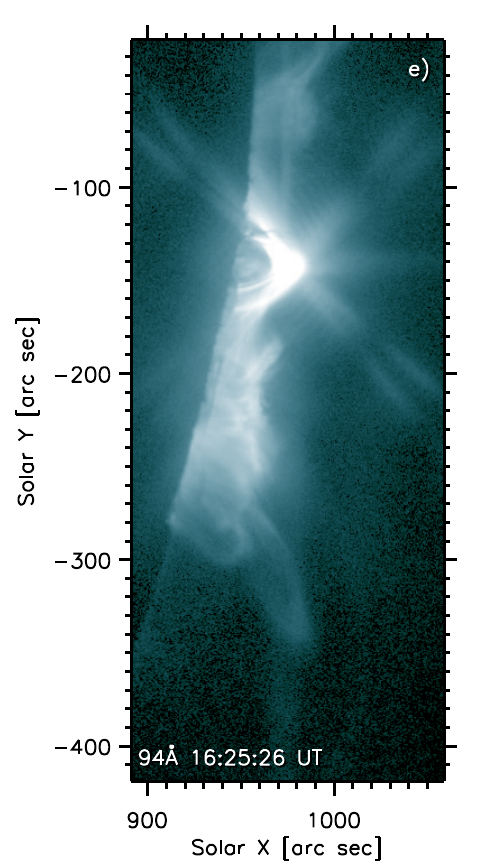}
   \includegraphics[width=3.53cm, clip,   viewport=  33 0 133 240]{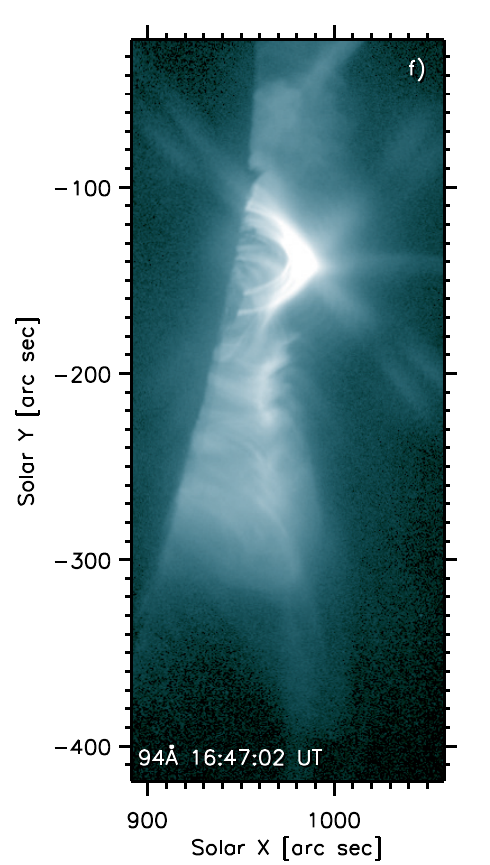}
   \includegraphics[width=3.53cm, clip,   viewport=  33 0 133 240]{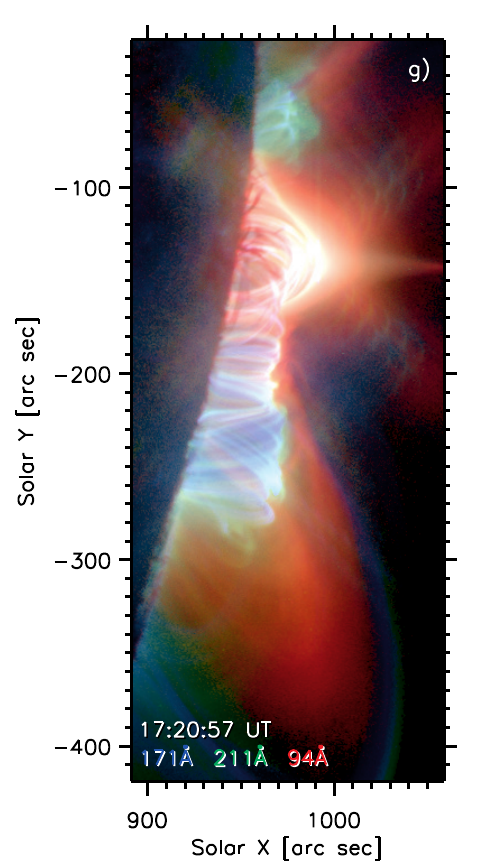}
  \caption{AIA observations of the saddle arcades during the 2017 September 6 (top row) and 2017 September 10 (bottom row) flares. Orange dashed lines in panel (a) indicate the flare ribbons. The filament as well as some of the loops composing the erupting flux rope of 2017 September 10 are indicated using arrows in panel (d). \label{fig_cpil_csheet}}
\end{figure*} 
\subsection{2011 December 26 flare and eruption}
\label{sec_20111226}

We next analyse a flare arcade which formed during the 2011 December 26 C5.7--class flare \citep{qiu17}. This flare occurred in the active region complex NOAA 11383 and 11384 and was accompanied by a Coronal Mass Ejection (CME) \citep{cheng16}.  

The flare morphology is shown in Figure \ref{fig_dudik_qiu}, panels (d)--(f). It showed a pair of very straight ribbons studied by \citet[][]{qiu17}. Nevertheless, we identified a pair of hooked extensions formed at their far ends, shown by arrows in panel (d). The western hook was curved and partially obscured by coronal loops rooted in the south-west at $\approx$[180$\arcsec$,360$\arcsec$], toward which the ribbon elongated.

First flare loops started to appear at around 11:15 UT in AIA 131\,\AA. In the composite image (panel (d)), they are seen as red-colored emission, located between the ribbons, and extending toward the western hook. As the arcade further developed (panel (e)), a saddle with two cantles started to be visible. The south-western cantle is more pronounced than the north-eastern one. This is because it was composed of long and faint flare loops, extending for more than 100$\arcsec$ toward the western hook in which they are rooted (arrow in panel (e)). On the other hand, the conjugate cantle started to be visible after the saddle cooled, at which it was visible in AIA 131 and 171\,\AA~(arrow in panel (f)).
\subsection{2017 September 6 eruptive flare}	
\label{sec_20170906}

On 2017 September 6, the large active region NOAA 12673 produced two flares \citep[e.g.,][]{yan18b}. The first one, confined X2.2--class one, started at $\approx$09:00 UT. Roughly 3 hours later, it was followed by a X9.3--class flare, the strongest one in the cycle 24 \citep[][]{peng20}, and accompanied by a CME. 

The eruption started at 11:55 UT. At this time, a pair of flare ribbons started to be visible in AIA 304\,\AA~data which we used for their tracing. The longer ribbon in the north-east developed a hook in the north, while the conjugate ribbon in the south-west contained a small hook towards the south (orange dashed lines in Figure \ref{fig_cpil_csheet}(a)). 

The X9.3 flare occurred in the gradual phase of the X2.2. As it started, it saturated the 131\,\AA~channel. Nevertheless, later on, a saddle can be readily identified at 12:35\,UT, see panel (a) of Figure \ref{fig_cpil_csheet}. Its northern cantle is composed of bright flare loops, observed to slip along the northern hook. The southern cantle is less bright. Still, some of the loops composing it are rooted near the southern hook. A fully-formed saddle is shown in panel (b). Similar to the events described before, the cooling saddle arcade is later visible for many hours in cooler filter channels, such as AIA 171\,\AA~and 211\,\AA, see panel (c).
\subsection{2017 September 10 eruptive flare}
\label{sec_20170910}

Four days later, when the same active region was behind the limb, it produced another large X8.2 flare \citep{yan18a,Warren18,Polito18}. The erupting flux rope consisted of a filament and a hot channel, studied in detail e.g., by \citet{yan18a}. The hot channel was visible in the AIA 94\,\AA~and 131\,\AA~and consisted of numerous loops located both to the north and to the south of the filament, see e.g., Figure 3 of \citet[][]{yan18a}, and \citet{chen20}.

The first flare loops can be identified at $\approx$16:00 UT. In panel (e) of Figure \ref{fig_cpil_csheet}, a saddle with cantles at both ends is readily identifiable. While the northern cantle is viewed along the axis of the saddle, the higher southern one is viewed from the side. This indicates that the flaring PIL was curved, which is in agreement with the observations of a curved filament residing in these locations before the eruption \citep{chen20}. Since the active region was located behind the limb, the flare ribbons could not be observed and it is impossible to relate the cantles with the hooks. However, location of the southern cantle corresponds to the previous southern extent of the erupting flux rope (southern arrow in panel (d)), while the northern cantle corresponds to the location of the filament eruption. 

The fully-developed saddle can be seen in panel (f) at 16:47 UT. It remained visible for more than 30 minutes (panel (g)). Even though the highest flare loops composing the southern cantle do not have a counterpart in cooler coronal channels at 17:20\,UT (panel f), the coronal channels still show that the central part of the flare arcade reaches lower altitudes than both its ends. 
\section{Summary}
\label{sec_summary}

We presented observations of five saddle-shaped arcades of flare loops formed during five well-known eruptive flares. The saddles are visible as the arcades develop a pair of cantles, longer, relatively-higher, and inclined flare loops located at both ends of the arcades. An example of a cantle loop observed in 171\,\AA~by both \textit{SDO}/AIA and \textit{Stereo}/EUVI showed that the cantle loops connect straight portion of one flare ribbon with the hooked extension of the conjugate ribbon. This suggests that the shapes of flare loops composing saddle-shaped arcades result from variations in magnetic connectivity.

The observational characteristics of the saddles are similar in all of the investigated eruptions. One peculiar difference concerns the development of the saddles during the 2011 June 7 and December 26 events (Sections \ref{sec_20110607}, \ref{sec_20111226}) where arcades initially formed a 'half-saddle' before the full saddle was seen. We suggest the formation of one cantle prior to its counterpart is caused by an asymmetric shape and/or evolution of the erupting flux rope, initially reconnecting with overlying arcades in a preferred direction.

Formation of flare loops joining the hooks and the straight parts of the conjugate ribbons is addressed in the latest extensions to the Standard flare model in 3D \citep[][and references therein]{aulanier19}. There, such loops form due to the ar--rf magnetic reconnection between the erupting flux rope and its overlying arcades rooted near the ribbon hooks. We found indications of this process using stereoscopic observations of the 2011 May 9 eruption, where it likely acted in the formation of one of the cantles. Furthermore, in the 2011 June 7 event, we identified footpoints of one cantle in the locations corresponding to footpoints of flare loops originating in the ar--rf reconnection reported earlier by \citet{dudik19}. Even though a complete analysis of the reconnection geometries in 3D was out of scope of this letter, these results still show that the shape of flare arcades could reflect the origin of individual flare loops. 

Finally, we shown that the flare arcades are saddle-shaped despite the fact that selected events were observed in different projections, magnetic environments, and accompanied by flares of different classes. This could indicate that the presence of saddles is common and a generic property of eruptive flares in 3D. 
\vspace{1 cm}
\\
The authors thank the anonymous referee whose comments helped us improve the manuscript. J.L. and J.D. acknowledge the project 20-07908S of the Grant Agency of Czech Republic as well as insitutional support RVO: 67985815 from the Czech Academy of Sciences. G.A. thanks the CNES and the Programme National Soleil Terre of the CNRS/INSU for financial support. AIA and data are provided courtesy of NASA/\textit{SDO} and the AIA science team. \textit{Hinode} is a Japanese mission developed and launched by ISAS/JAXA, with NAOJ as domestic partner and NASA and STFC (UK) as international partners. It is operated by these agencies in co-operation with ESA and the NSC (Norway). Full-disk EUVI images are supplied courtesy of the STEREO Sun Earth Connection Coronal and Heliospheric Investigation (SECCHI) team.

\textit{Facilities:} SDO, STEREO, Hinode


\end{document}